\documentstyle[12pt,graphicx,epsfig,graphics,epsf,psfig]{article}
\newcommand{\be}{\begin{eqnarray}}

\newcommand{\ee}{\end{eqnarray}}

\makeindex
\begin{document}
\pagestyle{plain}
\newcount\eLiNe\eLiNe=\inputlineno\advance\eLiNe by -1
\title{
         Long Range Forward-Backward Correlations and the Color Glass 
Condensate
}

\author{N\'estor Armesto$^1$, Larry McLerran$^2$ and Carlos Pajares$^1$ \\
\\
        $^1$ {\small \it Departamento de F\'{\i}sica de Part\'{\i}culas and
IGFAE,}\\
{\small \it 
Universidade de Santiago de Compostela,}\\
{\small \it
 15782 Santiago de Compostela,
         Spain} \\
       $^2$ {\small \it Nuclear Theory Group and Riken Brookhaven 
Research Center,}\\
            {\small \it Brookhaven National Laboratory}
{\small \it
        Upton, NY 11793  }
       }
 
\maketitle

\begin{abstract}
We discuss forward-backward correlations in the multiplicity of produced 
particles in heavy ion collisions.  We find the Color Glass Condensate
generates distinctive predictions for the long range component of this
correlation.  In particular, we predict the growth of the long range 
correlation with the centrality of the collision.  We argue that the
correlation for baryons is less strong than that for mesons.
\end{abstract}

\section{Introduction}

The Color Glass Condensate provides a phenomenologically successful
description of high energy hadronic processes~\cite{glr}-\cite{iv}.
The basis of the description for the 
infinite momentum hadronic wavefunction is to decompose the
wavefunction into pieces associated with the fast moving constituents,
and a piece associated with the slower moving components~\cite{rge}.
The fast moving 
components act as sources for the slow moving components.  The slow moving 
components are treated as classical fields.  Because the density of 
the gluons in the hadronic wavefunction is large, the coupling constant is 
weak.  The field strengths are large, nevertheless, $A \sim 1/g$, and although
one can use weak coupling techniques, the problem is essentially 
non-perturbative due to the strong field.

The physical picture of the soft component of the hadronic wavefunction
can be easily understood in terms of the phase space distribution
\be
	\eta = {{dN} \over {dyd^2p_Td^2r_T}}\,.
\ee
For $p_T < Q_{sat}$, where $Q_{sat}$ is the saturation momentum,
$\eta \sim 1/\alpha_s$.  This means that the quantum mechanical gluon states
are multiply occupied, and saturate at $1/\alpha_s$.  At this phase space
density, repulsive 
interactions become large.  As one adds more gluons to the system,
these components of the hadronic wavefunction remain 
fixed, since the repulsive interaction energy makes this unfavorable
relative to adding in a gluon in a less occupied state at higher transverse
momentum. Above the saturation momentum, the phase space density becomes 
small.  When one adds more gluons to the system, 
they typically have momenta above and near the 
saturation momentum.  This increases the
saturation momentum.  The renormalization group equations for the
Color Glass Condensate predict that the saturation momentum never
saturates, that is, it grows with decreasing $x$ of the gluons in the 
wavefunction~\cite{lipatov}-\cite{t}. One is perpetually 
adding in more gluons at an ever increasing saturation momentum.

When applied to hadron-hadron collisions, one imagines the collision
of two sheets of colored glass~\cite{Kovner:1995ja}-\cite{lappim}.
Initially, the CGC fields are transversely polarized color electric and 
magnetic fields.
At a very 
short time after the sheets collide, $t \sim e^{-\kappa/\alpha_s}$, the fields
become longitudinal electric and magnetic fields in the central region between
the colliding nuclei, and the
transverse field in this region vanishes.  The original transversely 
polarized fields are associated with the fast moving 
components of the nuclei and remain intact.
This situation is 
shown in Fig. 1.  The matter produced immediately after the collision
is called the Glasma. It has properties distinct from
the Color Glass Condensate and the Quark Gluon Plasma. It exist at times
intermediate between the two, and has some properties in common with both
the CGC and the QGP, hence the name.
\begin{figure}[h]
\begin{center}
\begin{tabular}{l l l}
\includegraphics[width=0.40\textwidth]{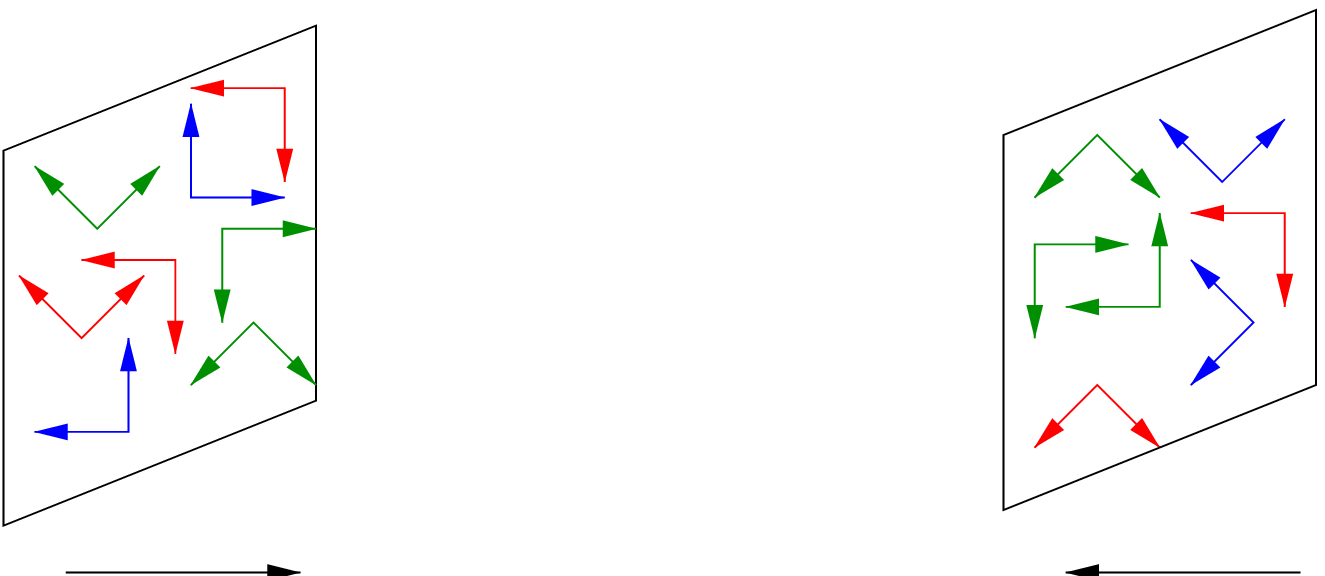} & &
\includegraphics[width=0.40\textwidth]{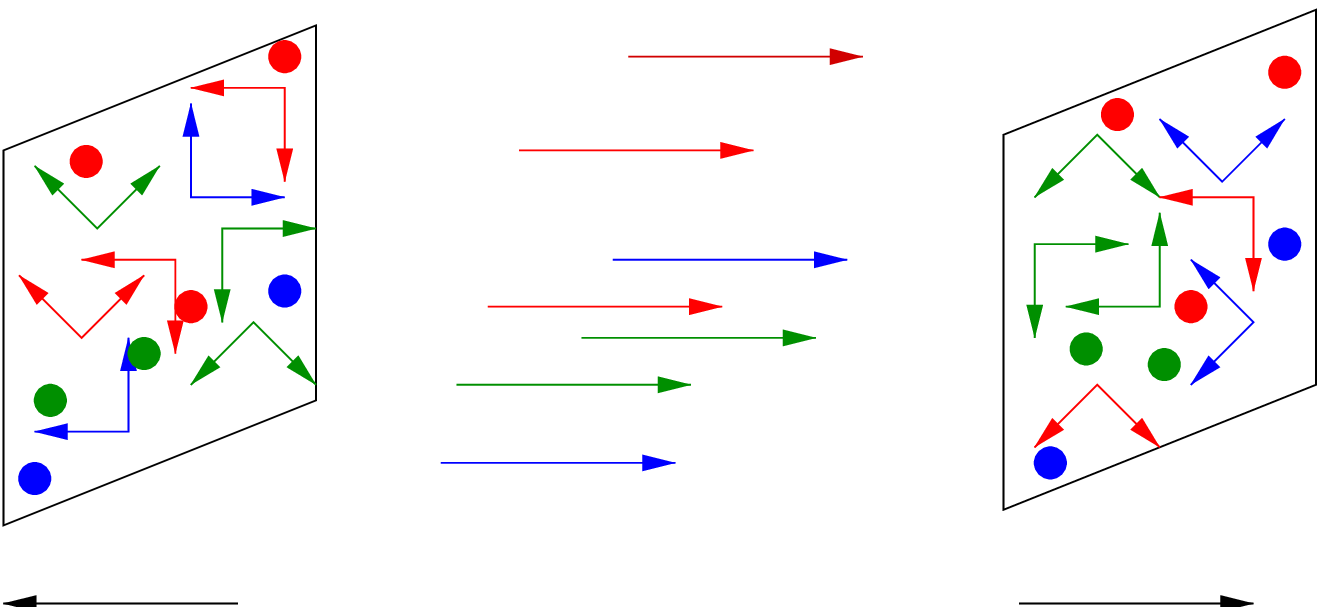}\\
 & & \\
 a & & b \\
\end{tabular}
\end{center}
\caption{ (a) Two sheets of colored glass approaching one another.
(b) After the collision, an equal and opposite charge density of color 
electric and color magnetic charges are set up on each nucleus.  This induces
longitudinal color electric and magnetic fields in the region between 
the nuclei.}
\label{collision}
\end{figure}

The origin of the longitudinal fields arises because as the sheets of colored
glass pass through one another, the fast moving components have added to them
a distribution of color electric and color magnetic charge.  The charge density
in the transverse plane of one nucleus is the negative of that in the
other nucleus. 

This picture of the fields has a restricted range of validity.  For  
the effective action which describes these fields to be valid, we require
that the field exist over a region of rapidity
\be
	\Delta y \le 1/\alpha_s\,.
\ee
In a larger range of rapidity, quantum fluctuations become important,
and the description in terms of classical fields breaks down.

It has also been argued 
that such long range
correlations exist in the dual parton string picture of hadron 
collisions~\cite{dpm,pajares}.
Some time ago, it was also argued that the Color 
Glass Condensate has long range
correlations in rapidity~\cite{klm}.
The Glasma provides a theoretical framework 
following directly from QCD which includes many features of the
dual parton model approach, in particular the longitudinal
rapidity structure.

In this paper, we explore these long range correlations.  We use the 
formalism of forward-backward correlations.  We argue that there is a long
range component of fluctuations for the gluons,
\be
	\left\langle {{dN^{gl}_F} \over {dy_1}}{{dN^{gl}_B} \over
{dy_2}}\right\rangle 
\sim {1 \over \alpha_s}
{{dN^{gl}} \over {dy}}\,,
\ee
which is approximately rapidity independent.  Here $F$ ($B$) are regions in
the positive (negative) rapidity region, and $y$ is some rapidity 
intermediate between $y_1$ and $y_2$, oftentimes taken to be
the center of mass rapidity.
For quarks,
\be
	\left\langle {{dN^{q}_F} \over {dy_1}}{{dN^{q}_B} \over
{dy_2}}\right\rangle
\sim 
{{dN^{q}} \over {dy}} \,.
\ee
The essential difference between the two is that for the gluons
the is a factor of $1/\alpha_s$.  This can be understood since the
gluons are acting coherently.  The decay process involves the coherent
emission of $1/\alpha_s$ gluons.  In the language of clusters, the 
typical cluster size of produced by a  fireball is $1/\alpha_s$ particles.
The quarks are fermions, and their typical cluster size is of order one,
since there is no coherence for fermions.

To compute the forward-backward correlation function, we also need to know
the short range component.  We define\footnote{This quantity coincides with
the $b$ usually defined through $\langle n_F\rangle (n_B) =a+b
n_B$~\cite{Capella:1978rg}.}
\be
        \sigma_{FB} = {{\left\langle N_F N_B\right\rangle -
\left\langle N_F\right\rangle\left\langle N_B\right\rangle} \over
{\left\langle N^2\right\rangle - \left\langle
N\right\rangle^2}}\,,
\ee
with $N$ the multiplicity in the forward or backward interval.
We argue below that there is 
a short range component which falls exponentially in rapidity.  It
is of strength of order $\alpha_s^2$ relative to that of the 
long range component.

We will argue below that at short range, the correlation function
is polluted by final state effects, and is not directly computed from
the initial state distribution of gluons. If we look at the long
range piece of $\sigma_{FB}$, say at rapidity separation of greater
that one unit of rapidity, final state effects should not be so important.

We argue below that this long range component has the form
\be
	\sigma_{FB} = {1 \over {1+c\alpha_s^2}}\,.
\ee
We do not perform a detailed computation of the constant $c$.
It involves knowing the strength of the soft correlated emission.
There is however a qualitative effect which we can understand about $c$:  
If the contribution of baryons
to the correlation function is large, then $c$ increases.  This
is because this contribution brings in no factor of $\alpha_s$.
The contributions of baryons
may be important of central collisions at large $p_T$.

Notice that as a function of increasing centrality, $\alpha_s$ decreases.
The forward-backward correlation function should therefore increase for 
more central collisions.

\section{Computing the Forward-Backward Correlation}

Consider the collision of two nuclei in the center of mass frame.  
We treat the high momentum degrees  
as sources.   There are sources for each nucleus.  Let us consider some 
region of transverse extent $a$.  The value of $a$ will be of the order of 
the inverse transverse momentum scale which we probe.  For the total 
multiplicity fluctuations, this size is of the order $a \sim 1/Q_{sat}$.

Now in the transverse region of size $a$, due to fluctuations, one of the 
nuclei will have a larger charge than the other.  The situation becomes
like that of $pA$ collisions where there is an asymmetry between target
and projectile~\cite{kovmu}-\cite{dumm}.  The total multiplicity
is given by the number of gluons in the object with the smaller charge,
evaluated at the saturation momentum scale of the larger charged object.
This is because the larger charge object is effectively a black disk for 
the smaller charged object, and as such, during the scattering 
the smaller charged object materializes all
of its gluons.
Since the difference in charge between these objects should be small
compared to the charge (so we can neglect the difference between the $Q_{sat}$
of
projectile and target), for each region of size $a$,
the total multiplicity is therefore of order
\be
	{{dN} \over {dy}} \sim {1 \over \alpha_s}~ a^2 Q_{sat}^2\,.
\ee
Summing over all areas gives the familiar Kharzeev-Nardi formula
for the total multiplicity~\cite{khnard},
\be
	{{dN} \over {dy}} \sim {1 \over \alpha_s} ~\pi R^2 Q_{sat}^2\,.
\ee

\begin{figure}[h]  
\begin{center}
\includegraphics[width=0.60\textwidth]{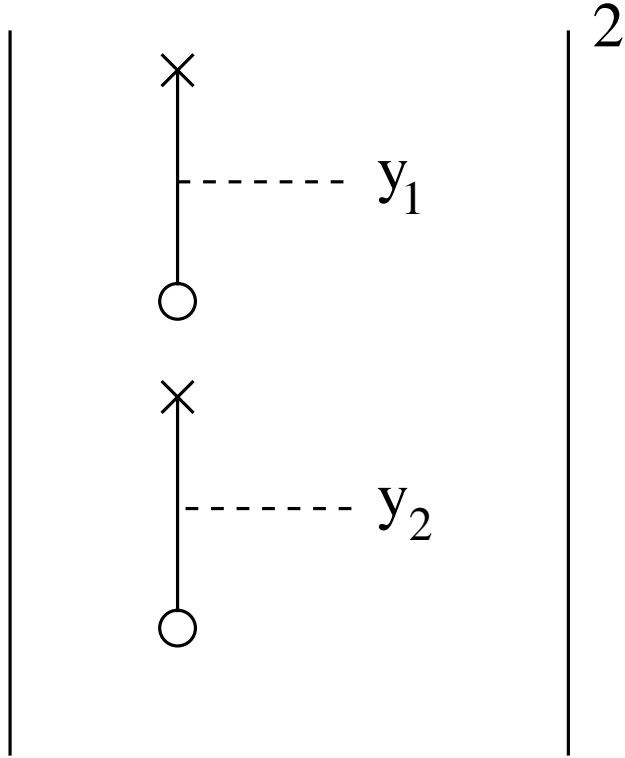}
\end{center}
\caption{ The leading order diagram which induce long range correlation
in rapidity.  The source of one nucleus is given by the $x$ and that of 
the other by the $o$.  The produced gluon is denoted by the dotted line.}
\label{clascor}
\end{figure}
The fluctuations are given by the square of the diagram for the multiplicity.
The
leading order contribution from the classical fields comes from the 
diagram of Fig. 2.  (This diagram at first sight appears to be uncorrelated,
but becomes correlated after averaging over the sources associated with the
nuclei.  This averaging ``ties together'' the sources in the upper and lower
part of the diagram.)
There is a factor of $1/\alpha_s^2$ which is due
to the fact that this is the square of two classical processes.  The
correlation will only be non-zero if the same transverse area is probed.
So the factor of $a^4$, upon summing over areas becomes of order $\pi R^2
a^2$.
There is also factor of $Q_{sat}^4$.  For the total multiplicity squared,
$a \sim 1/Q_{sat}$,
this gives
\be
	\left\langle \left({{dN} \over {dy}}\right)^2 \right\rangle 
\sim {1 \over \alpha_s^2} \pi R^2 Q_{sat}^2
\sim {1 \over \alpha_s} {{dN} \over {dy}}\,.
\ee
The fluctuation at different rapidities is the same since the field
which produces these particles is rapidity invariant:
\be
	\left\langle{{dN} \over {dy_1}} {{dN} \over {dy_2}}\right\rangle 
= \left\langle\left({{dN} \over {dy}}\right)^2\right\rangle\,.
\ee

Note the factor of $1/\alpha_s$ in this result.  This is because the 
$1/\alpha_s$ gluons associated with the classical field behave coherently.
Its origin is similar to that of the dependence on cluster size for 
fireball models of correlations.  One needs to know
the number of particles in each fireball decay.  This number
scales the square root in $N(\propto 1/\alpha_s)$
fluctuation since one assumes each
fireball is produced statistically but an extra factor of $N$ comes because
each fireball decays into $N$ particles.

There is of course a correlated piece, which comes form the diagram of Fig. 3.
\begin{figure}[h]  
\begin{center}
\includegraphics[width=0.60\textwidth]{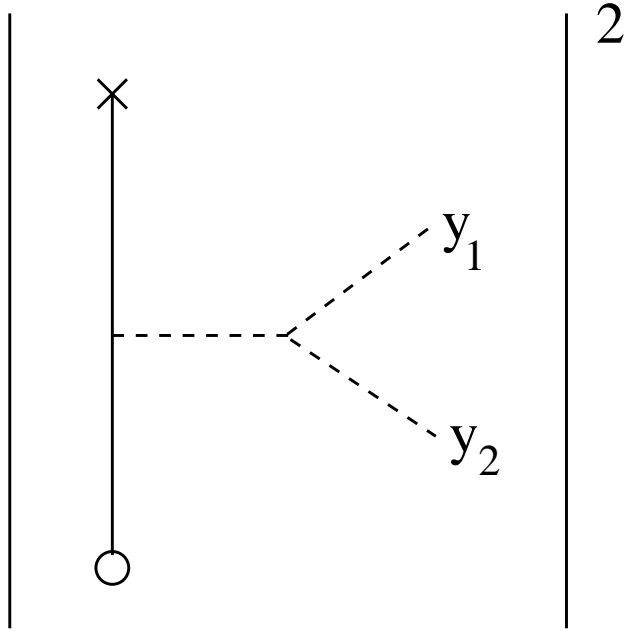}
\end{center}
\caption{ Here two gluon from the different sources scatter and produce
two gluons in the forward and backward hemisphere.}
\label{shortcor}
\end{figure}
Here one of the gluons is associated with the forward rapidity and the other 
with the backward rapidity.  This diagram has two factors of $\alpha_s$,
and should give a contribution to the total multiplicity fluctuations of order
\be
	\left\langle{{dN_{cor}} \over {dy_1}} {{dN_{cor}} \over {dy_2}}
\right\rangle \sim
 \alpha_s \pi R^2 Q_{sat}^4 \sim  {{dN_{cor}} \over
{dy}} 
e^{-\kappa \mid y_1 - y_2 \mid } \,.
\ee
Here $\kappa$ is a constant of order one.  
Note that $dN_{cor} /dy$ is of order $\alpha_s$ relative to that
for the classical contribution.  There is no interference with the leading 
order diagram of Fig. 2, since the average of an odd sources from the
same nucleus must vanish.  It should also be noted that outside of the
context of perturbative QCD, there should be diagrams like Fig. 3 associated 
with the scattering of the hadronic degrees of freedom of the nuclei.

To understand the different factors of $\alpha_s$ between the diagrams
of Fig 2 and Fig 3, we can think of the source as strong, and therefore 
of order $1/g$.  Therefore each of the diagrams squared in Fig 2 is of 
order $1/g^2$, making an overall contribution of order $1/\alpha_s^2$.
In Fig. 3, the diagram is overall zeroth order in powers of $g$.

For the long range correlation function, we see that the correlated piece
is overall a factor of $\alpha_s^2$ smaller than the classical contribution.
We therefore conclude that at large rapidities
\be
	\sigma_{FB} = {1 \over {1 + c \alpha_s^2}}\,,
\label{cent}
\ee
where $\alpha_s$ is to be evaluated at the saturation momentum associated 
the centrality of the collision and we have considered two intervals $F$, $B$
symmetric around midrapidity.  It clearly becomes more correlated
as the centrality increases.
It should also be noted that $c$ is an increasing function of the rapidity
difference $\Delta y = \mid y_1 - y_2 \mid$. To see it, we
define $\alpha_s \hat{c}= {{dN_{cor}} \over
{dy}} /{{dN} \over
{dy}}$ to get
\be
c=\hat{c} \,{1-e^{-\kappa \Delta y } \over 1+\alpha_s^2 \hat{c}
e^{-\kappa \Delta y
} }\,,
\ee
which clearly increases with increasing $\Delta y$.

We can repeat all the steps above for baryons.  The only difference is
that  the baryon multiplicity is of order $\alpha$ relative to that of
the gluons, and that in the fluctuation formula, there is no additional
factor of $1 /\alpha_s$ associated with coherent emission.  
Thus the baryon correlation
function should not grow with increasing centrality unlike of mesons
which presumably ultimately arise from gluons.

In the most central collisions at RHIC, at high $p_T$, there is considerable
contribution from baryons, and so measurements of correlations at high $p_T$ 
may not have such a large correlation for central collisions as 
would that for mesons.

In general, the correlations as a function of $p_T$ may be interesting.
The low $p_T$ gluons are in a maximally occupied state.  
The repulsive interaction of the gluon in these states will not allow the
addition of more gluons.  Changing the charges
on the nuclei therefore does not affect the multiplicity 
of low $p_T$ gluons, so there
should not be large correlation.  The dominant contribution to the total
multiplicity fluctuations comes from gluons with $p_T \sim Q_{sat}$.  For
$p_T > Q_{sat}$, there should be considerable correlation.  Of course the
transverse momenta of gluons is strongly affected by final state 
interactions, and it is unclear how this effect could persist.

As a final comment, let us recall the result for forward-backward
correlations in a different framework: string models which allow for some kind
of
collectivity~\cite{Amelin:1994mf}-\cite{Braun:2003fn} e.g. a
non-thermal phase transition like percolation. The generic finding in these
models is a reduction of the long range correlations with increasing
collectivity. While this reduction increases with increasing centrality, it
does not imply that the correlation itself reduces with increasing centrality
for a given modeling of collectivity.
Indeed, if $N_s$ is the number of independent
particle sources and $n_F$ the number of
particles in the forward hemisphere from each source,
the same assumptions leading to Eq.
(\ref{cent})
result here in
\be
\sigma_{FB}={1\over 1+A}\,,\ \ A={\langle N_s\rangle \over \langle N_s^2\rangle -
\langle N_s\rangle^2}\ { \langle n_F^2\rangle -\langle n_F\rangle^2 \over
\langle n_F\rangle^2}\,.
\ee
$A$ is~\cite{Cunqueiro:2005hx}
a decreasing function with increasing centrality, so
$\sigma_{FB}$ increases, as in the CGC, with increasing centrality.

\section{Summary and Conclusions}

The strongest objection to this analysis is that we have
used quarks and gluons as degrees of freedom.  In reality, we measure
mesons and baryons.  It is standard lore that the meson multiplicity 
distribution directly reflects that of the gluons.  Baryons, particularly
baryon-antibaryon pairs may be subject to more drastic final state 
interactions.  Baryons can be enhanced at intermediate $p_T$ in central
collisions.

For long range correlations, we have not imagined a final state
process which can affect the correlation in the total multiplicity.  
At short range, diffusion can
have a strong effect.  Imagine that after particles have been produced,
some of the particles from the forward region diffuse to the backward
region.  This decreases the number of forward particles and increase 
those in the backward direction.  This can induce a negative contribution
to the correlation function.  Certainly for rapidity differences of less 
than one unit, processes such as this will be important.  That said,
understanding such diffusion and comparison with correlation function
measurements may probe such effects, and may be interesting in their 
own right.

Fluctuations in the correlation between centrality and impact parameter may 
give long range correlations.  It is difficult to understand how the
correlation could become stronger with more central collisions.  Such
correlations can be estimated by a variety of Monte Carlo event generators,
and the strength of any resulting signal evaluated.

\section{Acknowledgments}

We thank the organizers of Quarks and Nuclear Physics 2006, where the
discussions concerning the work presented above were initiated.
We also thank Brijesh Srivastava for many enlightening conversations about how
the forward-backward correlations are measured in the STAR experiment.  
Larry McLerran thanks Dima Kharzeev for insightful critical comments,
and Rolf Scharenberg for useful insights.  This paper was based on data 
presented  at a  seminar given at Brookhaven National Laboratory, July 
27, 2006.\cite{bnl}

This manuscript has been authorized under Contract No. DE-AC02-98H10886 with
the U. S. Department of Energy.
NA is supported by Mi\-nis\-te\-rio de Educaci\'on
y Ciencia of Spain under a contract Ram\'on y Cajal. NA and CP acknowledge
financial support by CICYT of Spain under project FPA2005-01963.

\end{document}